          [2001/04/25 1.1 (PWD)]
\documentclass[a4paper,twocolumn]{esapub} 

\usepackage{times}
\usepackage{natbib}
\usepackage{graphics}

\title{Subsurface magnetic fields from helioseismology}
\author{H. M. Antia}
\affil{Tata Institute of Fundamental Research, Homi Bhabha Road,
Mumbai 400005, India}

\newcommand\be{\begin{equation}}
\newcommand\ee{\end{equation}}
\newcommand\jcd{Christensen-Dalsgaard}
\newlength{\figwidth}
\begin{document}


\maketitle

\begin{abstract}
Using even-order frequency splitting coefficients of global p-modes
it is possible to infer the magnetic field in the solar interior as
a function of radial distance and latitude. Results obtained using
GONG and MDI data are discussed. While there is some signal of
a possible magnetic field in the convection zone, there is little
evidence for any temporal variation of the magnetic field in the
solar interior.
Limits on possible magnetic field in the solar core are also
discussed.
It is generally believed that the
solar dynamo is located in the tachocline region.
Seismic studies do not show any significant temporal variation in
the tachocline region, though a significant latitudinal variation
in the properties of the tachocline are found. There is some evidence
to suggest that the latitudinal variation is not continuous and
the tachocline may consist of two parts.
\end{abstract}

\section{Introduction}

Helioseismology has been successful in probing the spherically
symmetric structure (Gough et al.~1996) of the Sun as well as the rotation
rate in its interior (Thompson et al.~1996; Schou et al.~1998).
To the first order, rotation affects only the frequency splitting coefficients
which represent odd terms in the azimuthal order $m$ of the oscillation
modes. The even terms in these splitting coefficients, can arise from
second order effects of rotation, magnetic field or any latitudinal
dependence in the structure.
It is not possible to distinguish between the effects of
a magnetic field and aspherical perturbations to the solar structure
(Zweibel \& Gough 1995).

The even order splitting coefficients are fairly small, and no
definitive results have so far been obtained regarding the
magnetic field strength in the solar interior. Dziembowski \&
Goode (1989) using data from the Big Bear Solar Observatory
claimed to find evidence for a mega Gauss field near the base of
the convection zone. Improved data from the Global Oscillation
Network Group (GONG) project (Hill et al.~1996) and the
Michelson Doppler Imager (MDI) instrument (Rhodes et al.~1997)
on board the
SOHO satellite has not confirmed these results
(Antia et al.~2000).

Instead of a magnetic field one can invoke aspherical structure to
explain the even coefficients of frequency splittings. In this
case, it is possible to apply an inversion technique to determine
the latitudinal dependence in solar structure variables like the
sound speed and density (Antia et al.~2001a).
The advantage of this approach is that
it can give the location of perturbation giving rise to the
observed even splitting coefficients. 

The GONG and MDI instruments have been observing the Sun for the
last 7 years and it is also possible to study possible temporal
variation in the internal magnetic field. It is well known
that the frequencies of solar oscillations vary with time and this
variation is correlated with solar activity (Elsworth et al.~1990;
Libbrecht \& Woodard 1990). Similarly, the even splitting coefficients
are also known to vary with time and their variation is correlated
to the corresponding component of observed magnetic flux at the solar
surface (Libbrecht \& Woodard 1990; Woodard \& Libbrecht 1993;
Howe et al.~1999; Antia et al.~2001a).
However, most of these temporal variations are found to arise
from perturbation near the solar surface (Basu \& Antia 2000; Antia
et al.~2001a). There is little evidence for any significant
temporal variations in the solar structure below the thin surface
layers.

\begin{figure*}
\hbox to \hsize{\resizebox{\figwidth}{!}{\includegraphics{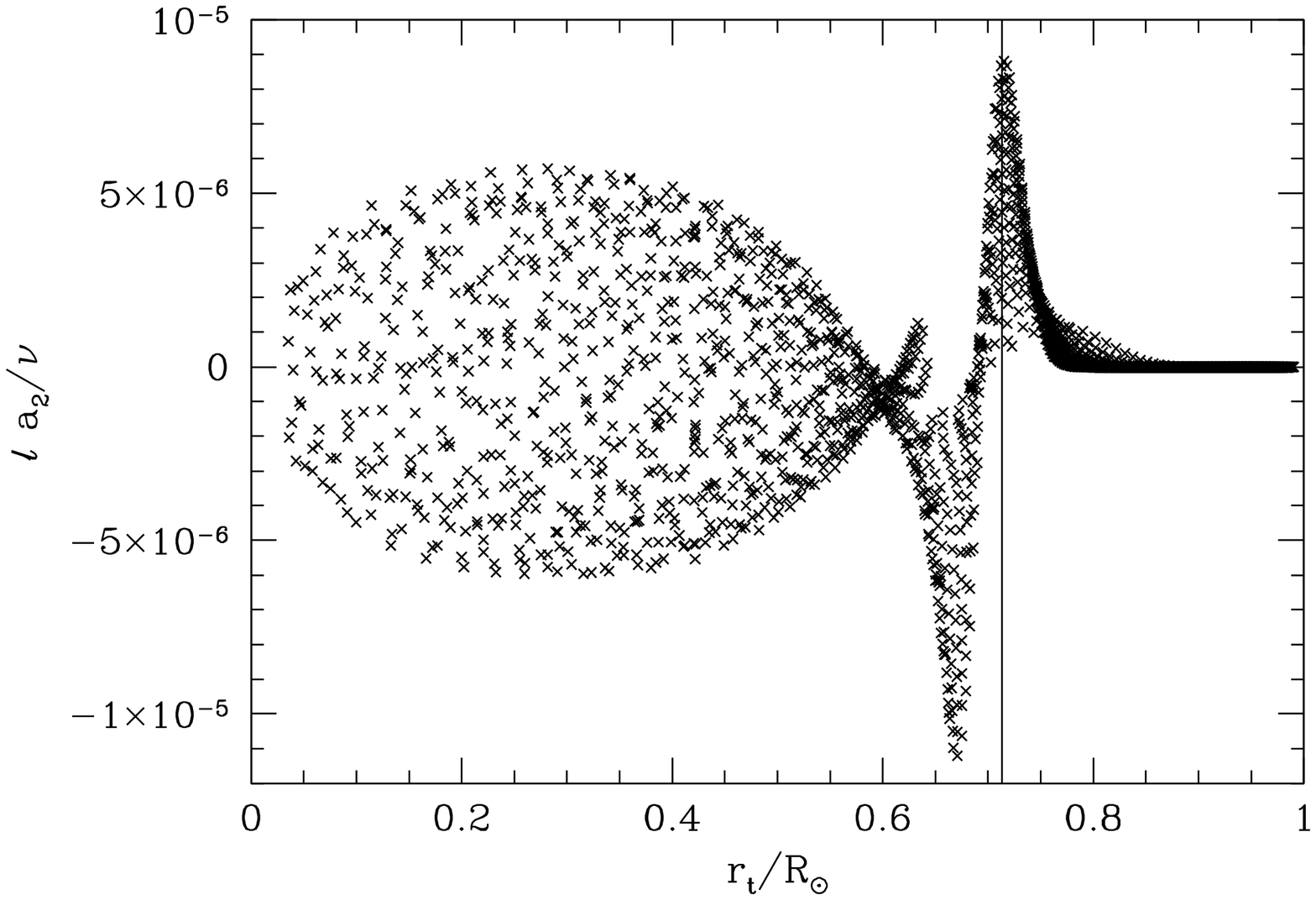}}
\hfil\resizebox{\figwidth}{!}{\includegraphics{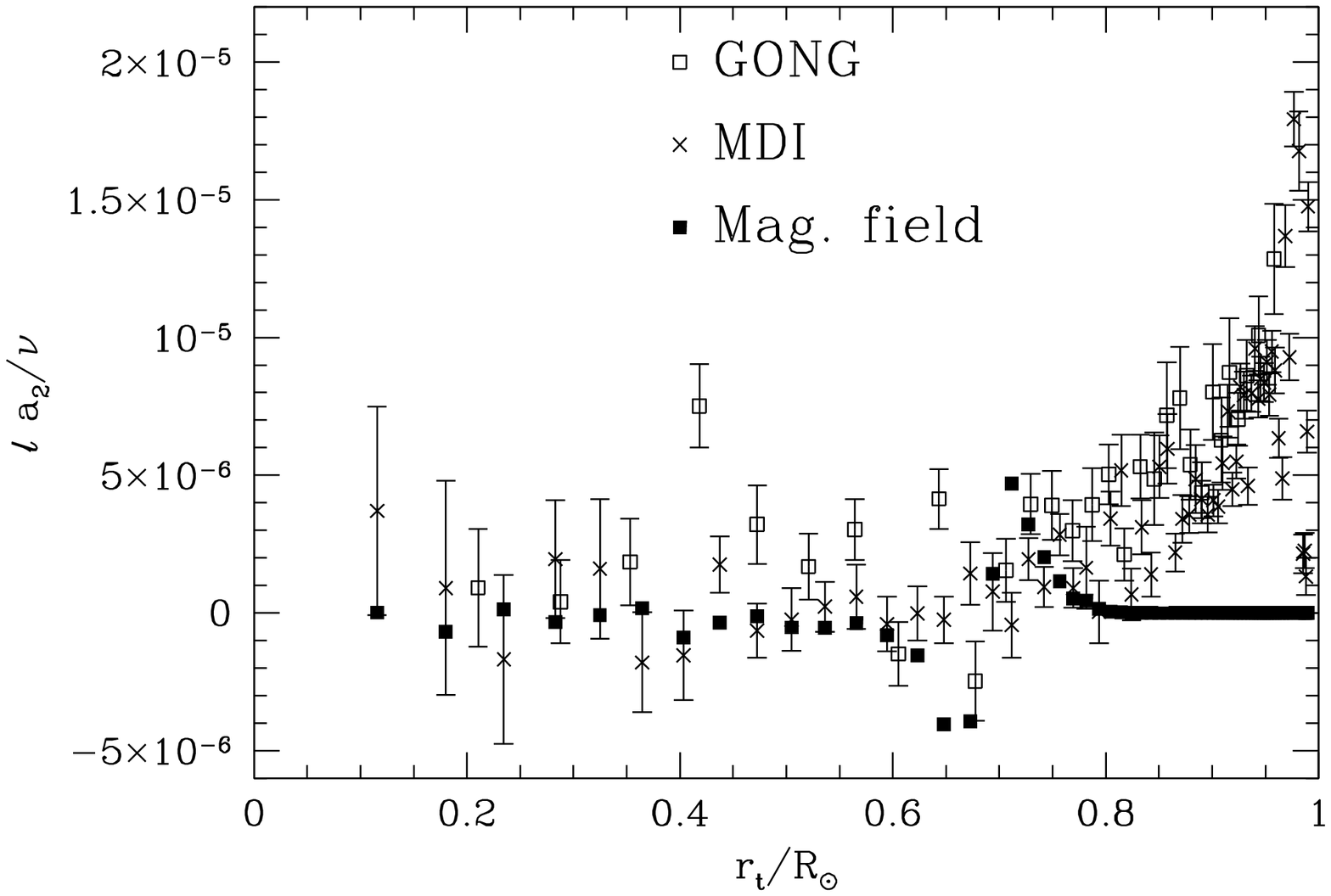}}}
\caption{
The left panel shows the splitting coefficients $a_2$ from
a toroidal magnetic field concentrated near the base of the
convection zone, plotted
as a function of the lower turning point of the modes.
The magnetic field is given by Eqs.~(2,3) with $k=2$, $\beta_0=10^{-4}$,
$r_0=0.713R_\odot$ and $d=0.02R_\odot$. In the right panel
these coefficients are compared with observed values.
Each point represents an average over 25 neighbouring modes.
The estimated contribution from rotation has been subtracted
from the observed splittings plotted in the figure.}
\end{figure*}

Inversions for rotation rate (Thompson et al.~1996; Schou et al.~1998)
have shown that the observed differential rotation at the solar
surface continues through the convection zone, while in the radiative
interior the rotation rate is more or less independent of latitude.
The transition takes place close to the base of the convection zone
in a region which has been named as the tachocline (Spiegel \& Zahn 1992).
It is generally believed that the solar dynamo operates in the
tachocline region. Hence it would be interesting to look at the
temporal variations in the solar structure and rotation rate in
the tachocline region. Howe et al.~(2000) found a 1.3 year
periodicity in the equatorial rotation rate at $r=0.72R_\odot$.
But other investigations (Antia \& Basu 2000; Corbard et al.~2001)
did not find any systematic variation in the same region.
Helioseismic inversions are unreliable in the tachocline region
and the properties of the tachocline have been studied using forward
modelling approach (Kosovichev 1996; Antia et al.~1998; Charbonneau
et al.~1999). These results have also not shown any significant
temporal variations in the tachocline properties (Basu \& Antia 2001).

Although no temporal variations have been seen in the tachocline
properties, there is a definite latitudinal variation in the position
and possibly also in the thickness of the tachocline (Charbonneau et
al.~1999; Basu \& Antia 2001). On the other hand, there is no
latitudinal variation in the depth of the convection zone or
the solar structure in the tachocline region. These results
appear to be contradictory as it is believed that the tachocline
region is mixed by some rotationally induced instability
(Richard et al.~1996; Brun et al.~1999). Hence, it would be interesting
to study the latitudinal variation in the tachocline with accumulated
data over the last seven years.

The global modes of oscillations used in these studies can only give
information about the large scale structure and magnetic field in the
solar interior. To study smaller features like active regions we need
to use local helioseismic techniques, like the time-distance
analysis (Duvall et al.~1993) or the ring diagram technique (Hill 1988).
In this work we present some results obtained using the ring
diagram technique, while the time-distance
analysis of active regions is described by Kosovichev (2002).

\section{Effect of a magnetic field on solar oscillations frequencies}

In the absence of rotation or magnetic field the frequencies of
solar oscillations are independent of the azimuthal order $m$.
Rotation or magnetic field break this degeneracy leading to splitting
of frequencies with the same radial order $n$ and degree $\ell$.
The frequencies of solar oscillations can be expressed in terms of the
splitting coefficients:
\be
\nu_{n,\ell,m}=\nu_{n,\ell}+\sum_{j=1}^{J_{\rm max}} a_j^{n,\ell}
{\cal P}_j^{\ell}(m),\qquad (J_{\rm max}\le 2\ell)
\label{split}
\ee
where ${\cal P}_j^\ell(m)$ are orthogonal polynomials of degree $j$ in $m$
(Ritzwoller \& Lavely 1991; Schou et al.~1994).
The odd coefficients $a_1, a_3,a_5,\ldots$,
can be used to infer the rotation rate in the solar interior, while the
even coefficients arise basically from second order effects due to rotation and
magnetic field. Since forces due to rotation or magnetic field in the
solar interior are smaller by about 5 orders of magnitude
as compared to the gravitational forces, it is
possible to apply a perturbative treatment to calculate their
contribution to frequency splittings (Gough \& Thompson 1990).
Since the rotation rate in the solar interior can be inferred from
the odd splitting coefficients, we can use this inferred profile to
calculate the expected contributions to the even splitting coefficients
from the second order effects of rotation. This calculated contribution
can be subtracted out from the observed splitting coefficients to get
the residuals which may be due to a magnetic field or other asphericities.

For simplicity we consider only a toroidal magnetic field, taken
to be of the form,
\be
{\bf B}=\left[0,0,a(r){dP_k\over d\theta}{(\cos\theta)\atop
\phantom{d\theta}}\right],
\label{mag}
\ee
with the axis of symmetry coinciding with the rotation axis.
Here, $P_k(x)$ is the Legendre polynomial of degree $k$ and
\be
a(r)=\cases{\sqrt{8\pi p_0\beta_0}(1-({r-r_0\over d})^2)
&if $|r-r_0|\le d$\cr
0 &otherwise\cr}
\label{magtor}
\ee
where $p_0$ is the gas pressure, $\beta_0$ is a constant
giving the ratio of magnetic to gas pressure,
$r_0$ and $d$ are constants defining
the mean position and half-thickness of layer where the field is
concentrated.

\subsection{The seismic data}

We use data sets from both GONG and MDI for this study.
The solar oscillations frequencies and the splitting coefficients
from the GONG project (Hill et al.~1996) were obtained from 108 day time
 series. The
MDI data were obtained from 72-day time series (Schou 1999).
These data sets consists of the mean frequency and the splitting
coefficients for each $n,\ell$ multiplet. We use the 62 data sets
from GONG covering a period from May 7, 1995 to July 21, 2001.
Each set covers data from 108 days with a spacing of 36 days between
consecutive data sets. The MDI data sets (Schou 1999) consist of
28 non-overlapping sets covering a period from May 1, 1996 to
March 30, 2002, with a break between July 1998 and January 1999
when the contact with SOHO was lost.

\subsection{Field near the base of the convection zone}

Since there have been some suggestions that a toroidal magnetic
field may be concentrated near the base of the
convection zone (Dziembowski \& Goode 1989) we consider
splitting coefficients arising from such a field with
$r_0=0.713R_\odot$, $d=0.02R_\odot$, $\beta_0=10^{-4}$ and $k=2$.
Figure 1 shows
the resulting splitting coefficients $a_2$ as a function
of the lower turning point of the modes. These coefficients show
a characteristic signature which should be detectable in the
observed splittings if a strong enough magnetic field is
indeed present in these layers. The errors in observed splitting
coefficients is too large to show this signal and hence we take
averages over neighbouring modes to reduce error bars.
The resulting $a_2$ after subtracting out the contribution
due to rotation is shown in the right panel of Fig.~1,
which also shows the expected
signal averaged in the same manner for a magnetic field with $\beta_0=10^{-4}$.
The MDI data is from the first year of observation and the GONG data
is from averaged spectrum for GONG months 4--14. Both these data sets
are from observations around the minimum in solar activity.
From this figure it can be seen that there is no clear signature
of any feature near the base of the convection zone in the
observed splittings. Thus we can only set an upper limit on the
magnetic field in this layer, which will of course, depend on the
thickness of the magnetic layer. For a layer with half-thickness of 
$0.02R_\odot$ the upper limit turns out to be $\beta_0=7\times10^{-5}$
or a magnetic field strength of 300 kG near the base of the convection
zone (Antia et al.~2000). Similar limits have been obtained by
Basu~(1997). This limit is consistent
with the value independently inferred by D'Silva \& Choudhuri (1993).

\subsection{Field in the upper convection zone}

Looking at Fig.~1 it appear that there is no signature of a magnetic
field in the radiative interior, as the observed splitting coefficients fall
off smoothly with depth of the lower turning point.
The only noticeable feature in the observed
splitting coefficients is the peak around $r=0.96R_\odot$. If this
peak is solely due to a magnetic field, the field may be distributed
around this depth. This is approximately the depth to which the
outer shear layer in rotation profile extends (Antia et al.~1998;
Schou et al.~1998). Comparison with expected splittings from magnetic
field indicates that the observed peak may be due to a field with
$\beta_0=10^{-4}$ (or $B\approx 20$ kG) concentrated around
$r=0.96R_\odot$ (Antia et al.~2000). 

We do not expect an ordered large scale magnetic field inside the
convection zone. It is possible that there is a concentration of
randomised magnetic field at this depth which gives rise to the
observed peak in the splitting coefficients. The splitting due to
such a field will be quite different from what we have calculated
and hence it is difficult to get the exact form or strength of
the magnetic field from these calculations. Alternately, it is
possible that this peak is due to departure from spherical symmetry
in solar structure. This question is discussed in the next section.
This feature is found to be present in all data sets and does not
appear to depend significantly on the solar activity level.

\section{Asphericity in the solar structure}

\begin{figure}
\resizebox{\figwidth}{!}{\includegraphics{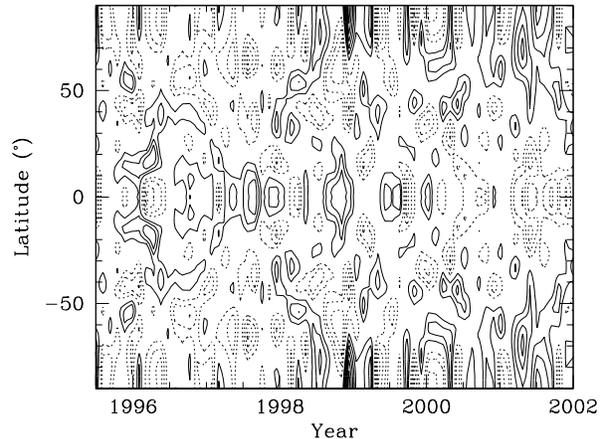}}
\caption{
Contour diagram of the  aspherical component of squared sound speed
at $r=0.96R_\odot$ as a function of time and latitude using the
GONG data. The solid
contours correspond to positive values, while dotted ones are for
negative values. The contour spacing is $2\times10^{-5}$. The figure
shows the residual obtained after subtracting out the temporal
average at each latitude.}
\end{figure}

\begin{figure*}[th]
\hbox to \hsize{\resizebox{\figwidth}{!}{\includegraphics{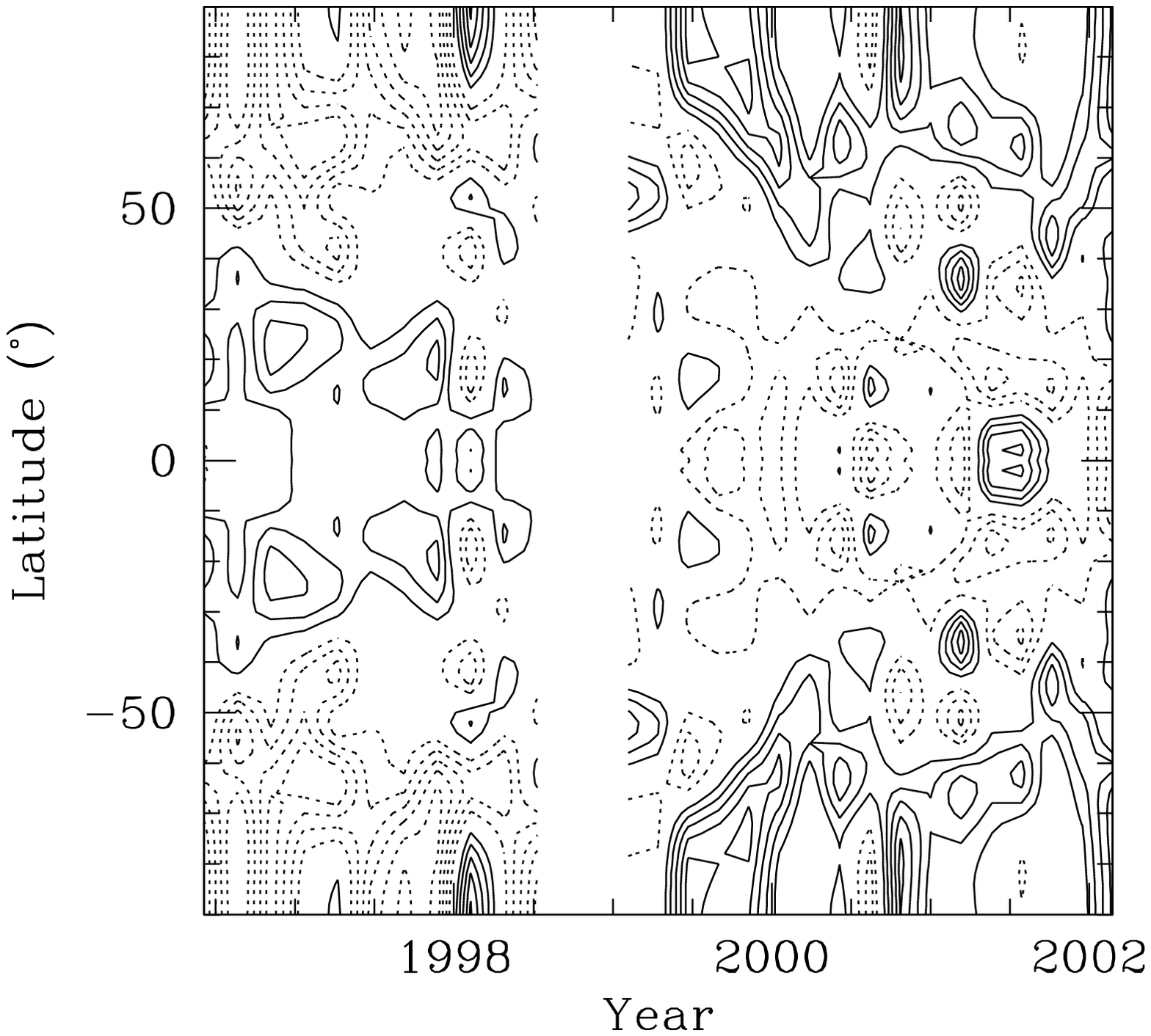}}
\hfil\resizebox{\figwidth}{!}{\includegraphics{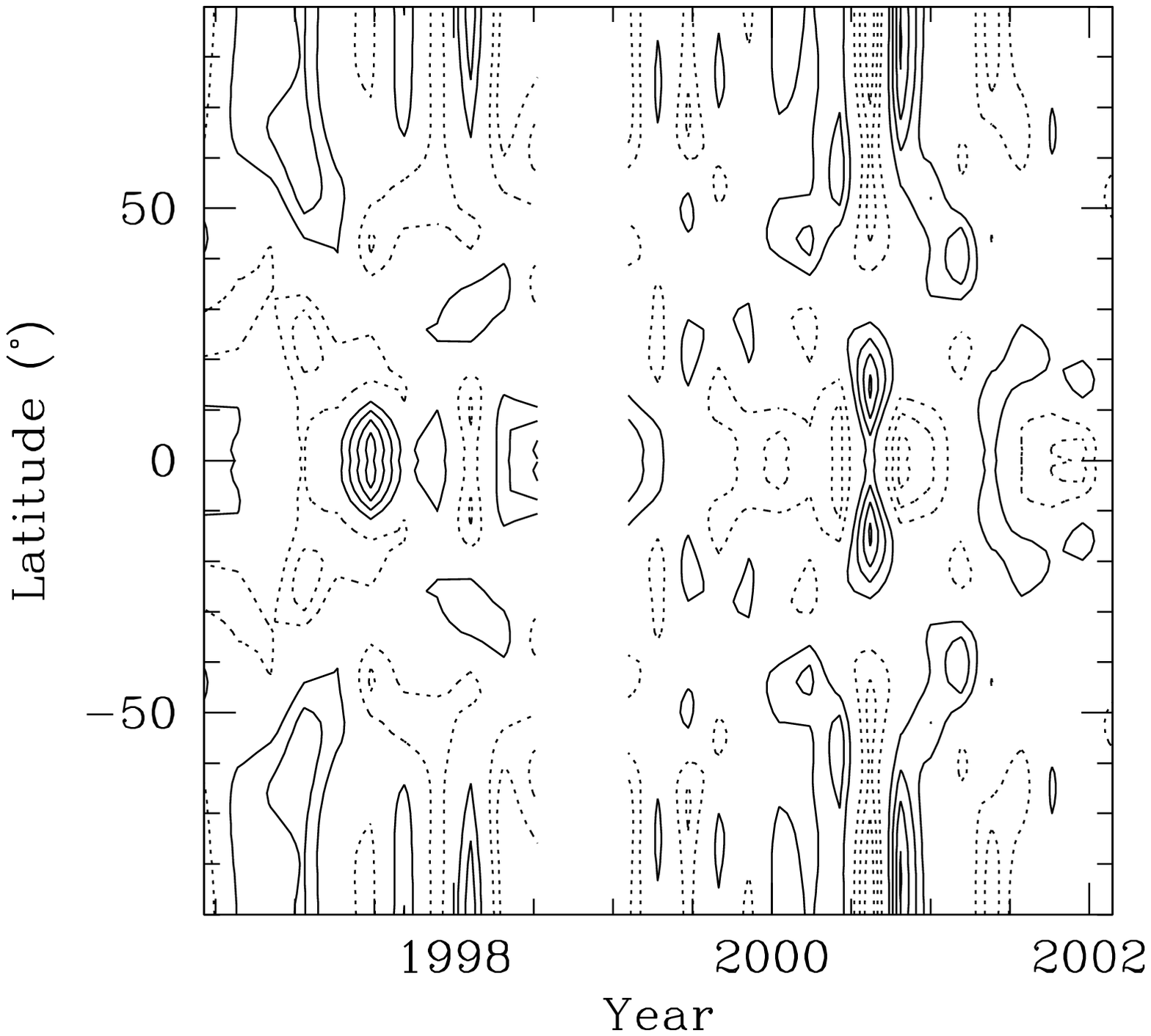}}}
\caption{
Contour diagram of the aspherical component of squared sound speed
at $r=0.96R_\odot$ as a function of time and latitude using the MDI data.
This figure
shows the residual obtained by subtracting out the temporal average
at each latitude.  The solid
contours correspond to positive values, while dotted ones are for
negative values. The contour spacing is $2\times10^{-5}$. The left
panel shows the results obtained using all modes with $1.5\le\nu\le3.5$
mHz, while the right
panel shows the results using modes with $1.5\le\nu\le3.0$ and
$\ell<110$.}
\end{figure*}

Apart from second order contributions from rotation, which can be
easily estimated, the even splitting coefficients can arise from
either a magnetic field or departures from spherical symmetry in solar
structure. We have discussed the first possibility in the previous
section. In this section we consider the other possibility. The
departure from spherical symmetry can arise due to magnetic field
(in which case this is included in the calculations described in the
previous section) or may arise because the Sun may not be in strict
hydrostatic equilibrium as is the case inside the convection zone.
There could be other contributions to pressure, e.g., turbulent
pressure, which are not isotropic and these
may cause departure from spherical symmetry. Leaving aside
the origin of asphericity we attempt to generalise the structure
inversion technique to include departures from spherical symmetry.
The splitting coefficients are sensitive
only to the north-south symmetric component of asphericity and hence
only this component can be determined. The differences in the sound
speed, $c$, and the density,
$\rho$, with respect to a spherically symmetric solar
model can be calculated using
(Antia et al.~2001a)
\begin{eqnarray}
\lefteqn{{\ell a_{2k}{(n,\ell)}\over\nu_{n\ell}}=
Q_{\ell k}{F_k(\nu_{n\ell})\over E_{n\ell}}+
{Q_{\ell k}(4k+1)\over2}\times
} \label{eq:inv}\\
& &\hspace{-0.7 cm}\int_0^R\;dr\;\int_0^\pi\sin\theta\;d\theta
\left({\cal K}_{c^2,\rho}^{n\ell}
{\delta c^2\over c^2}+{\cal K}_{\rho,c^2}^{n\ell}{\delta \rho\over \rho}
\right)P_{2k}(\cos\theta)\nonumber
\end{eqnarray}
where $E_{n\ell}$ is the mode inertia (\jcd\ \& Berthomieu 1991) and
$Q_{\ell k}$ is a geometric factor as defined by Antia et al.~(2001a).
Here $F_k(\nu)$ are the surface terms which accounts for uncertainties
in the treatment of surface layers.

\subsection{Temporal variations in asphericity}

Eq.~\ref{eq:inv} can be used for inversion to determine the sound speed
and density as a function of radial distance and latitude.
Using seismic data at different times it is
also possible to study temporal variations in the solar structure.
It is well known that the frequencies and even order splitting
coefficients vary with solar activity. Thus it would be interesting
to study if this implies any variations in the solar structure.
The spherically symmetric component has been well studied and will not be
considered in this work.

Fig.~2 shows the results for aspherical component of squared sound speed
obtained using GONG data at $r=0.96R_\odot$
as a function of time and latitude.
To see the temporal variations
more clearly, the figure shows the residuals
obtained after subtracting the temporal average at each latitude.
There is no pattern in the residuals, thus suggesting that there
is no significant temporal variation. Similar results are obtained
at other depths also.
The aspherical component of density also shows a similar result and
does not show any significant temporal variation.
Thus it is clear that all the temporal variation
in the splitting coefficients is accounted for by that in the surface
term. In fact the surface term is well correlated with the corresponding
component of the observed magnetic flux at the solar surface
(Antia et al.~2001a).

Fig.~3 shows the residuals in sound speed at $r=0.96R_\odot$ as
determined by the MDI data. The left panel which shows the results
using full data shows a distinct temporal variation in contrast
to the GONG results. However, a careful look at the figure suggests
that most of the temporal variation occurs between July 1998 and
January 1999, when the contact with SOHO satellite was lost. This
period is
marked by the gap in the contour diagram. Before the gap
there is positive asphericity at low latitudes and negative values
at high latitudes. This is reversed after the gap and in fact there
is very little systematic variations in the results before or after
the gap. Thus there are two possibilities, either the Sun has
had some interesting transition exactly when the MDI instrument
was not operational, or the apparent variation is due to systematic
errors introduced due to instrumental variations during recovery
of the SOHO satellite. Considering the fact that the GONG data does not
show any abnormality during this period, the second possibility
appears to be more likely. This seems to be confirmed by the
right panel which shows the same results but using only modes
with $\ell<110$. In this case there is no temporal variation in
asphericity. Thus it appears that the systematic errors in MDI data are
predominantly in the high degree modes.

\begin{figure*}
\hbox to \hsize{\qquad\resizebox{0.92\figwidth}{!}{\includegraphics{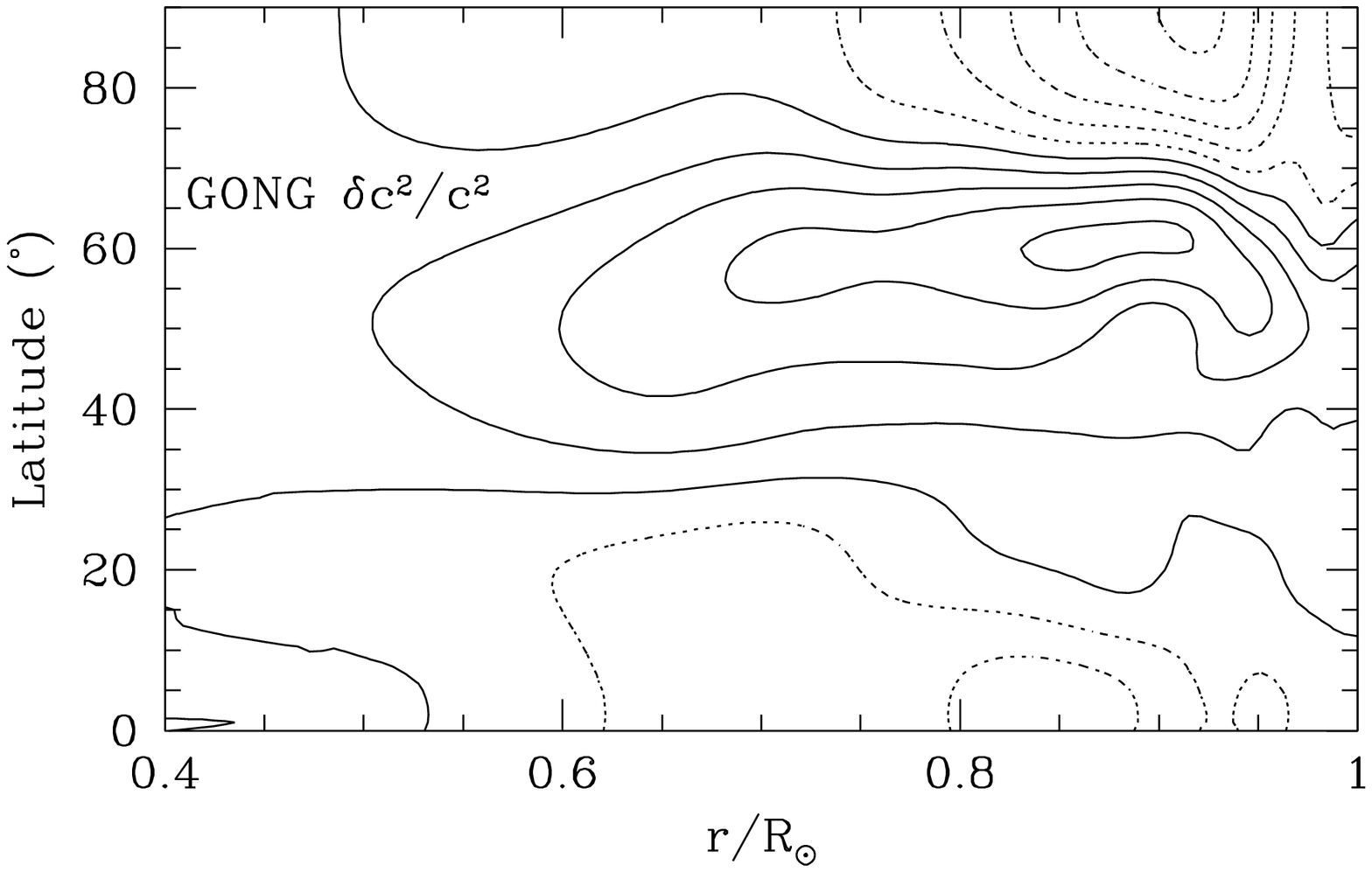}}
\hfil\resizebox{0.92\figwidth}{!}{\includegraphics{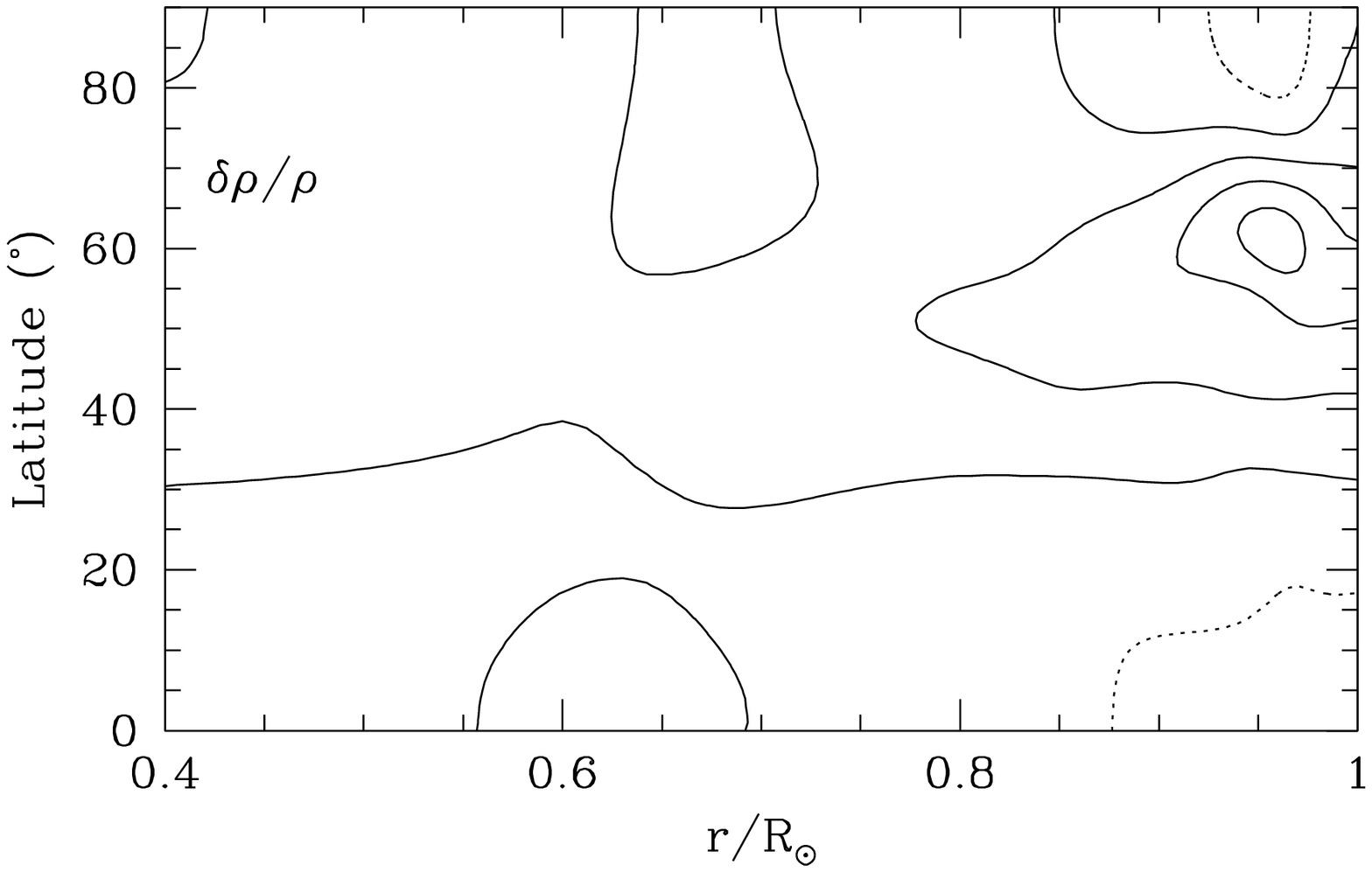}}\qquad}
\bigskip
\hbox to \hsize{\qquad\resizebox{0.92\figwidth}{!}{\includegraphics{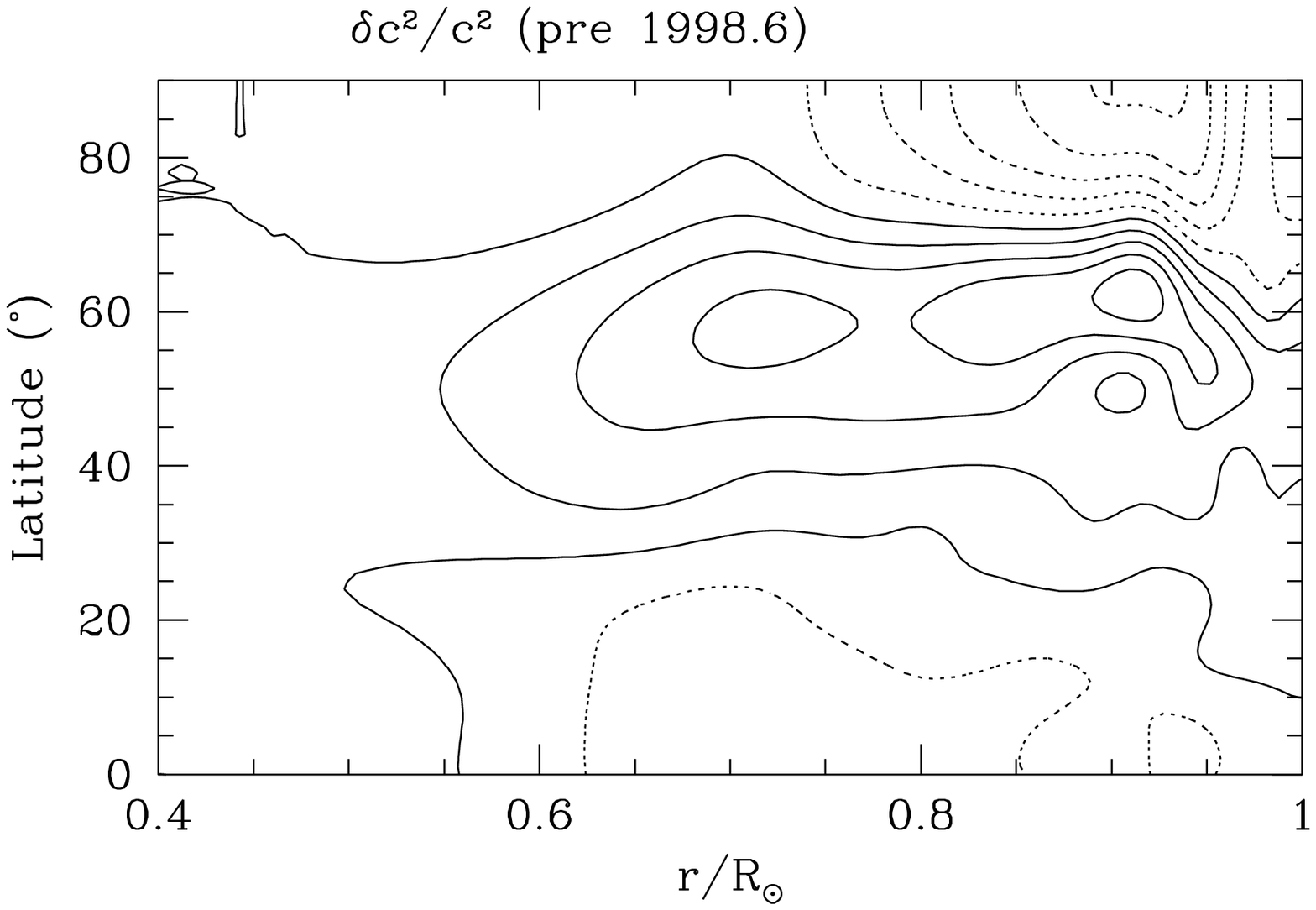}}
\hfil\resizebox{0.92\figwidth}{!}{\includegraphics{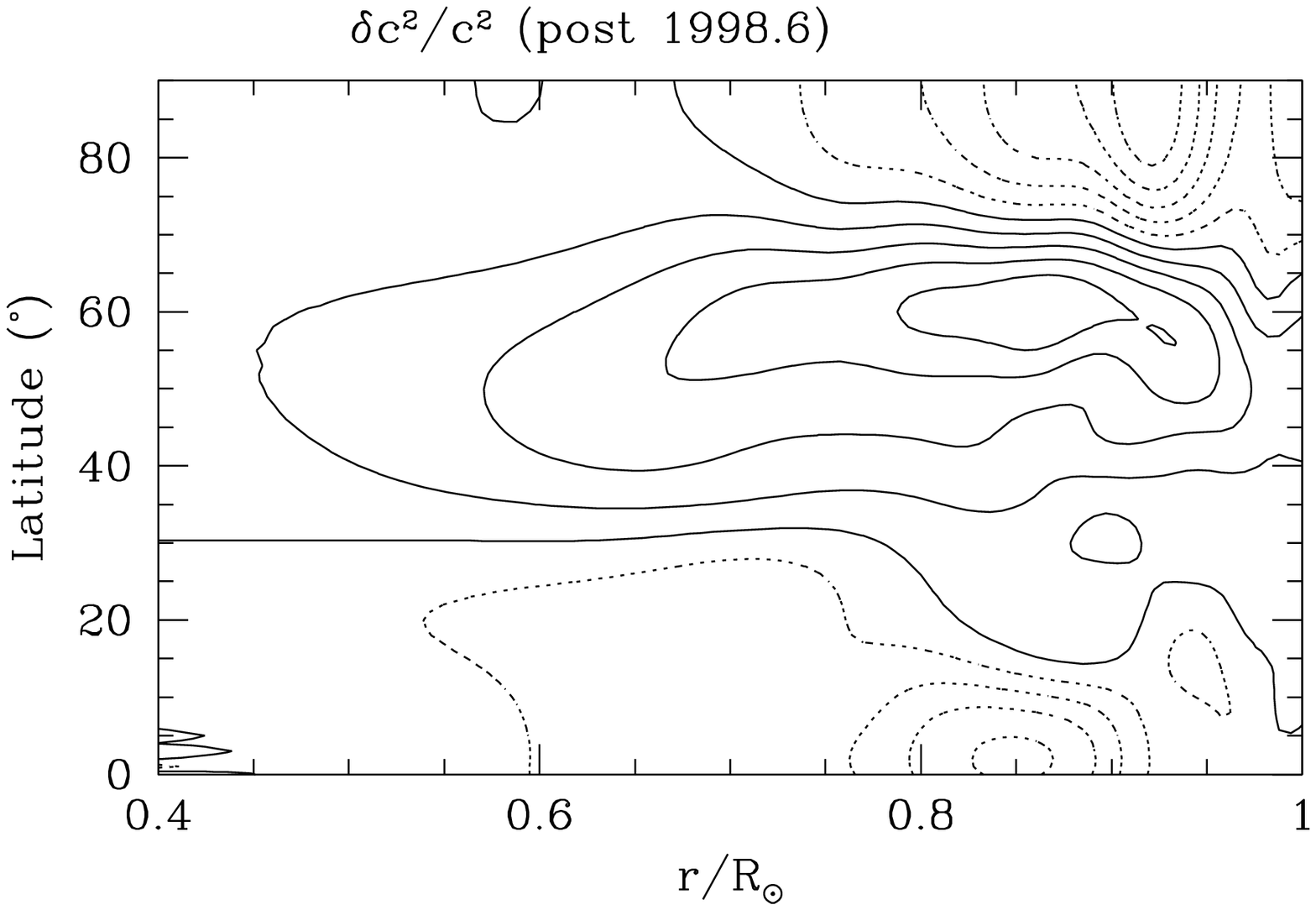}}\qquad}
\caption{
Contour diagram of the temporal average of aspherical component of
squared sound speed (top left panel) and density (top right panel)
as a function of radial distance and latitude using the GONG data.
The solid
contours correspond to positive values, while dotted ones are for
negative values. The contour spacing is $2\times10^{-5}$.
The lower panels show the temporal average over pre 1998.6 and post
1998.6 data.}
\end{figure*}

These systematic errors in MDI data are likely to affect other
conclusions too. For example, from the observed variation in
the solar f-mode frequencies Dziembowski et al.~(2001) have
concluded that the solar radius is reducing at the rate of
1.5 km per year. Looking at their Fig.~2 (results with $\gamma_f$ included)
it appears that
most of the variation has occurred between 1998.5 and 1999.2,
the same period when MDI was not operational. There is very little
variation in estimated radius before or after the break. This becomes
more clear if additional data that has now become available is
added. Thus
once again the claimed radius variation is probably an effect of
systematic errors. In fact, using an independent analysis
Antia et al.~(2001b) did not find any significant variation in the
solar radius. 

\subsection{Temporal average of asphericity}

Since we do not find any significant temporal variation in
asphericity in the solar interior, it is possible to improve the
accuracy of inversions by taking temporal average over all data sets.
The average asphericity from GONG data as a function of radius
and latitude is shown in Fig.~4. The MDI results obtained using
only modes with $\ell<110$ are similar to this. The typical errors
in low latitude region inside the convection zone is about $10^{-5}$
which is half the contour spacing. The errors increase with depth
and latitude. The most striking feature is the broad peak around
$r=0.9R_\odot$ and a latitude of $60^\circ$. This is similar to
the feature seen in section 2.3 in the splitting coefficients.
The peak has shifted downwards when the surface term
is removed for inversion.
The maximum asphericity in sound speed is $10^{-4}$.
The perturbation in density is significantly smaller
than that in squared sound speed. This may imply that a major
part of this perturbation is coming from a magnetic field. A
thermal perturbation will affect the density and squared sound
speed (which is essentially temperature) to a comparable extent.
Of course, this is not essential and there could be thermal
perturbation which don't affect the density significantly.
As noted
earlier this feature could be due to a random magnetic field in the
convection zone. If that is the case we can argue that the propagation
speed is altered due to magnetic field and we may expect
$\delta c^2/c^2\approx v_A^2/c^2$,
where $v_A$ is the Alfven speed. This will translate to
a field of about 70 kG at $r=0.9R_\odot$.

Near the base of the convection zone $\delta c^2/c^2\approx5\times10^{-5}$
around a latitude of $60^\circ$. If this is due to a magnetic field the
field strength would be about 250 kG, comparable to the upper limit
obtained in Section 2.2. It should be noted that in Section 2.2 we
only considered field concentrated near the base of the convection
zone, while the feature seen in Fig.~4 is spread over a broad
region. Thus the earlier limit does not apply to this. From
Fig.~4 if we try to put a limit on the magnetic field concentrated near
the tachocline, then it will be of the order of
$v_A^2/c^2\sim 2\times10^{-5}$ (twice the estimated error in $\delta c^2/c^2$),
which will correspond to a field strength of
about 150 kG, close to the value inferred by D'Silva \& Choudhuri
(1993). Thus it is not clear if any significant field as required by
dynamo theories is concentrated in this region.

To study the possibility of small temporal variation in inferred
asphericity in sound speed, we divide the data sets into two parts
around middle of 1998 and take averages separately over these parts.
The results are shown in Fig.~4. There is a distinct increase in
asphericity in post 1998.6 set as compared to earlier times. The
increase appears to be up to $2\times 10^{-5}$ in some regions and is
spread over a wide region.

\section{Magnetic field in the solar core}

The inversions for asphericity become unreliable in the solar core
($r<0.3R_\odot$) since very few p-modes penetrate into this region.
Further, the few modes that penetrate have large errors in the
splitting coefficients and since the sound travel time in the
core is fairly small the splitting coefficients are not very
sensitive to the magnetic field in this region.
As a result, it is difficult to get much information
about possible magnetic field or asphericity in the solar core
from these even order splitting coefficients. However, most of the
solar mass is in the core and any significant perturbation to
spherical symmetry in that region will cause significant distortion
which will be visible even at the surface, unless it is compensated
by suitably large distortions in the outer layers. From the
temporal average of asphericity shown in Fig.~4, we can conclude
that for $r>0.3R_\odot$, $\delta c^2/c^2<10^{-4}$ except possibly close
to the peak in asphericity around $r=0.9R_\odot$ and latitude of
$60^\circ$. It would be interesting to examine if this limit
is applicable to possible asphericity in the solar core also.

\begin{table}
  \begin{center}
    \caption{Distortion at solar surface due to a toroidal magnetic
field located at different layers in the solar interior.}
      \vspace{1em}
    \renewcommand{\arraystretch}{1.2}
    \begin{tabular}[h]{ccc}
      \hline
      $r_0/R_\odot$ & $d/R_\odot$ & $|\Delta r/r|$ \\
      \hline
0.2&0.1&$2\times10^{-4}$\\
0.4&0.1&$6\times10^{-5}$\\
0.6&0.1&$1\times10^{-5}$\\
0.8&0.1&$6\times10^{-6}$\\
      \hline \\
      \end{tabular}
    \label{tab:tab1}
  \end{center}
\end{table}

Using the formalism developed by Gough \& Thompson (1990) it is
possible to calculate the distortion due to a large scale magnetic
field in the solar interior. If we assume that the magnetic field
is toroidal and given by Eqs.~2,3, with $\beta_0=10^{-4}$ and
$k=2$ we can calculate the resulting distortion at the solar surface
and the results for a few different values of $r_0$ are summarised
in Table 1.
It is clear that a
magnetic field in the core produces distortion at the solar surface
which is comparable to $\beta_0$, while if the field is located
in the outer regions, the distortion is much less. This may be
expected because the fractional mass affected by the magnetic field is
small when the field is in outer regions.

The expected distortion at the solar surface can be compared with
the observed distortion of $-(5.4\pm0.5)\times10^{-6}$ 
(Kuhn et al.~1998). Of this the seismically inferred
rotation profile accounts for a distortion of $-5.8\times10^{-6}$
(Antia et al.~2000).
Thus the residual distortion at the surface
is at a level of $10^{-6}$. This will imply that the magnetic field
in the core should be such that $\beta_0<10^{-6}$.
The actual
distortion will depend on the form of the magnetic field and thickness
of region where it is located. To allow for all these factors
we can increase the upper limit by an order of magnitude. Thus
we can put a conservative upper limit of $10^{-5}$ on the ratio of
magnetic to gas pressure in the solar core.

\section{The tachocline}

Using the forward modelling technique described by Antia et al.~(1998)
we study the properties of the tachocline as a function of time using
the GONG and MDI data. The rotation rate in the tachocline is modelled
by (Antia et al.~1998)
\be
\Omega_{\rm tac}=
{\delta\Omega\over {1+\exp[(r_t-r)/w]}},
\ee
where
$\delta\Omega$ is the jump in the rotation rate across the tachocline,
$w$ is the half-width of the transition layer, and $r_t$ the
radial distance of the mid-point of the transition region.
No significant temporal variation is found in any of these properties
of the tachocline (Basu \& Antia 2001).
Thus we can take temporal average to improve the accuracy and the
results are shown in Table 2.

\begin{table}
  \begin{center}
    \caption{Properties of the tachocline at a few selected latitudes.}
      \vspace{1em}
    \renewcommand{\arraystretch}{1.2}
    \begin{tabular}[h]{lccc}
      \hline
$\!\!$Lat.&$\delta\Omega$&$r_t$&$w$\\
($^\circ$)&(nHz)&($R_\odot$)&($R_\odot$)\\
      \hline
\phantom{0}0&$\phantom{-}20.8\pm0.3$&$.6917\pm.0023$&$.0062\pm.0012$\\
15&$\phantom{-}17.8\pm0.2$&$.6910\pm.0021$&$.0076\pm.0010$\\
45&$-30.6\pm0.4$&$.7097\pm.0021$&$.0103\pm.0012$\\
60&$-67.8\pm0.6$&$.7104\pm.0027$&$.0151\pm.0015$\\
      \hline \\
      \end{tabular}
    \label{tab:tab2}
  \end{center}
\end{table}

From Table~2 it is clear that there is significant latitudinal
variation in the position and thickness of the tachocline, which
is consistent with the earlier results (Charbonneau et al.~1999;
Basu \& Antia 2001; Corbard et al.~2001). However, from the
table it appears that the latitudinal variation is not continuous.
The position and thickness for latitudes of $0^\circ$ and $15^\circ$
are the same within error-bars. Similarly, those for latitudes
of $45^\circ$ and $60^\circ$ are close to each other, but they are
significantly different from the values for low latitudes.
Thus it appears that the tachocline actually consists of two parts,
one at low latitudes where $\delta\Omega>0$ and another at high
latitudes where $\delta\Omega<0$. This two parts are located at
different depths and have different thicknesses, but there may be
no significant latitudinal variation within each part. Thus
the tachocline may cover the shaded region in Fig.~5, where we have
used a half-thickness of $2.5w$, which would cover about 85\% of the
variation in the rotation rate.

\begin{figure}
\centerline{\resizebox{\figwidth}{!}{\includegraphics{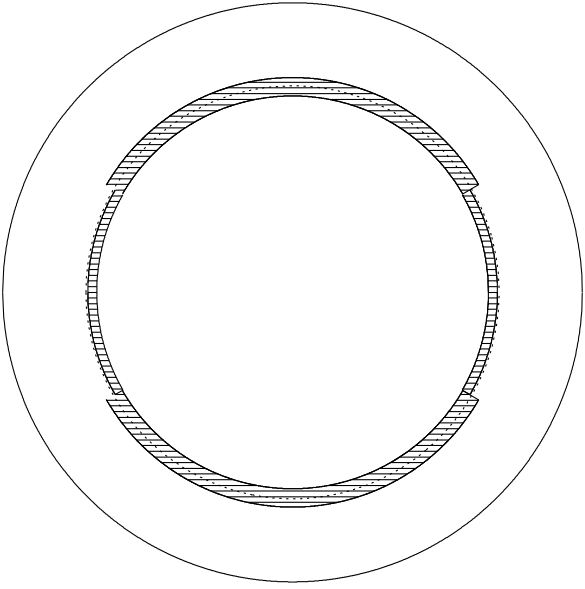}}}
\caption{
A section of the Sun showing the position of the tachocline (shaded area) and the base
of the convection zone (dashed line). The solar rotation axis is
assumed to be vertical.}
\end{figure}

It is believed that large shear in the tachocline region gives
rise to some instabilities which cause mixing in the tachocline
region. This mixing accounts for low observed lithium abundance
at the solar surface and also yields sound speed profile in
better agreement with seismic inversions (Richard et al.~1996;
Brun et al.~1999). If the tachocline has a significant latitudinal
dependence, then we would expect the mixed region also has
a latitudinal dependence giving rise to asphericity in solar
structure. However, there is no evidence for significant asphericity
in this region as seen from Fig.~4 and further the depth of the
convection zone also does not have any significant latitudinal
variation (Basu \& Antia 2001). This will appear to suggest that
the tachocline may not be responsible for the mixing below the
base of the convection zone. Another alternative is that the
thickness of the tachocline varies in such a manner that the
lower boundary of the mixed region does not have significant
latitudinal dependence. This appears to be the case in Fig.~5.
Of course, we have chosen the width of $2.5w$ to achieve this.
Nevertheless, this is a reasonable estimate for extent of mixing
due to the tachocline. It may be argued that it is a coincidence
that variation in position and thickness match each other.
If there is no mixing in the tachocline,
as in the standard solar model (\jcd\ et al.~1996), the deviation
in squared sound speed is of order of 0.5\%, while the asphericity
in the tachocline region is of order of $5\times10^{-5}$. Thus
the agreement between position and thickness variations in the
tachocline should be at level of 1\%. A part
of asphericity seen inside the convection zone may be due to
latitudinal variation in the tachocline.

\section{Magnetic field in active regions}

To study the subsurface magnetic field in active regions we use
the ring diagram technique (Hill 1988) which
involves study of
3d spectra from observations over a part of the solar disk.
We use the available spectra from the MDI data for this study.
Each spectra covers a region of $15^\circ$ in heliographic
longitude and latitude. Each spectra is fitted following
the procedure explained by Basu et al.~(1999) and Basu \&
Antia (1999) to calculate the mode parameters. To study the
influence of active region on mode properties, we choose a
pair consisting of a quiet and an active region at the same
latitude. The difference in mode properties between these two
regions will give the effect of magnetic field in the active region
(Rajaguru et al.~2001). It is found that frequencies increase with
surface magnetic field. This frequency difference cannot be
accounted by the surface term, which shows that the difference
between active and quiet region persists to a depth of at least
few Mm. Further, the frequency shift in f-modes is comparable to
those in p-modes of similar frequency, which probably implies that
the difference is arising from a magnetic field rather than thermal
perturbation. The thermal perturbations are not expected to affect
the f-modes significantly.
But it is difficult to infer the form or location
of the field from the seismic data. It is possible that Wilson
depression associated with the Sunspot can account for the frequency
differences between active and a quiet regions.
It may be noted that because of spatial resolution of $15^\circ$ on
the solar surface we can only study average properties over regions
much larger than sunspots.
Time-distance analysis
(Kosovichev 2002) is better suited to study the magnetic field
in active region because of higher spatial resolution.

\section{Summary}

Using the even order splitting coefficients it is possible to study
the magnetic field and departures from spherical symmetry in the
solar interior. Unfortunately, it is not possible to distinguish
between these two possibilities. The seismic data from GONG and MDI
covering the last 7 years does not show any signal from possible
toroidal magnetic field concentrated near the base of the
convection zone. An upper limit on such a concentrated field
is about 150 kG.
The seismic data shows a broad feature around
$r=0.9R_\odot$ and a latitude of $60^\circ$ which may be due to
a magnetic field or aspherical perturbation to the solar structure.
The aspherical perturbations to the sound speed are at the level
of $10^{-4}$ in this region and if these are due to a magnetic field
we may expect a field strength of 70 kG. We do not expect large
scale ordered magnetic field inside the convection zone but it
is possible that some randomised magnetic field is present.
This feature extends to the base of the convection zone where it
has a magnitude of $5\times10^{-5}$, which will correspond to a
field strength of 250 kG. The observed splitting coefficients
do not yield a tight constraint on the magnetic field in the solar
core as very few modes penetrate to the core. However, any magnetic
field in this region will yield significant distortion at the solar
surface. From the observed distortion we can put an upper limit of
$B^2/(8\pi p_0)<10^{-5}$ in the solar core ($r<0.4R_\odot$).
This corresponds to a field strength of 7 MG at the centre, or
3MG at $r\approx0.2R_\odot$ or 0.8 MG at $r\approx 0.4R_\odot$.

There is no significant temporal variation in aspherical component
of sound speed or density in the solar interior. This also applies
to possible temporal variations in magnetic field. Thus any
temporal variation in solar interior is less than about
$5\times10^{-5}$, which is the error estimate inside the convection
zone at low latitude. By taking temporal average over the high
activity and low activity data sets it appears that there
may be a small increase in sound speed asphericity with activity
at the level of $10^{-5}$ in a broad region centred at
latitude of $60^\circ$. This is comparable to the error estimate
and its significance is not clear.
The MDI data do
show some temporal variation, but a closer look shows that most of
the temporal variation has occurred during the time when SOHO had
lost contact. Thus this is likely to be an artifact of systematic
error introduced during recovery of SOHO. This systematic error
will also affect other inferences about temporal variation obtained
using the MDI data. This systematic error appears to be predominantly
in high degree ($\ell>110$) modes.

It is generally believed that the solar dynamo is operating in the
tachocline region. But no significant temporal variation is seen in
properties of the tachocline. Nevertheless, the tachocline is known to be
prolate and there is indeed a significant latitudinal variation in
the position and thickness of the tachocline. However, this latitudinal
variation may not be continuous and it appears
that the tachocline may consist of two parts one at low latitude
where the rotation rate increases with radius and second one at
high latitude where the rotation rate decreases with radius. These
two parts may be located at different depths and have different
thicknesses. But there may be no significant variations within each
part. The thickness difference between the two parts should match the depth
variation to ensure that the lower limit of mixing due to
the tachocline is essentially independent of latitude.

\section*{Acknowledgements}

This work  utilises data obtained by the Global Oscillation
Network Group (GONG) project, managed by the National Solar Observatory,
which is
operated by AURA, Inc. under a cooperative agreement with the
National Science Foundation. The data were acquired by instruments
operated by the Big Bear Solar Observatory, High Altitude Observatory,
Learmonth Solar Observatory, Udaipur Solar Observatory, Instituto de
Astrofisico de Canarias, and Cerro Tololo Inter-American Observatory.
This work also utilises data from the Solar Oscillations
Investigation/ Michelson Doppler Imager (SOI/MDI) on the Solar
and Heliospheric Observatory (SOHO).  SOHO is a project of
international cooperation between ESA and NASA.


\end{document}